\documentclass{iaus}
\usepackage{graphicx}

\title[Cold Dust Properties of Early-Type Galaxies]{Using GAMA and H-ATLAS Data to Explore the Cold Dust Properties of Early-Type Galaxies}
{}

\author[Nicola K. Agius]   
{Nicola K. Agius$^1$, Anne E. Sansom$^1$, Cristina C. Popescu$^1$}

\affiliation{$^1$Jeremiah Horrocks Institute, University of Central Lancashire, \\ Preston, 
PR1 2HE, Lancashire, United Kingdom  \\ email: {\tt NKAgius@uclan.ac.uk}}

\pubyear{2011} 
\volume{284}  
\pagerange{1--12}
\jname{The Spectral Energy Distribution of Galaxies}
\editors{R.J. Tuffs \&  C.C.Popescu, eds.}
\begin{document}

\maketitle

\begin{abstract}
Hierarchical galaxy formation models predict the development of elliptical galaxies through a combination of the mergers and interactions of smaller galaxies. We are carrying out a study of Early-Type Galaxies (ETGs) using GAMA multi-wavelength and Herschel-ATLAS sub-mm data to understand their intrinsic dust properties. The dust in some ETGs may be a relic of past interactions and mergers of galaxies, or may be produced within the galaxies themselves. With this large dataset we will probe the properties of the dust and its relation to host galaxy properties. This paper presents our criteria for selecting ETGs and explores the usefulness of proxies for their morphology, including optical colour, Sersic index and Concentration index. We find that a combination of criteria including r band Concentration index, ellipticity and apparent sizes is needed to select a robust sample. Optical and sub-mm parameter diagnostics are examined for the selected ETG sample, and the sub-mm data are fitted with modified Planck functions giving initial estimates for the cold dust temperatures and masses.

\end{abstract}

\firstsection 
\section{Introduction}

Early-Type Galaxies (ETGs) are thought to be a homogeneous class of E and S0 galaxies (\cite[Baldry \textit{et al.} 2004]{Baldry2004}). They are predominantly old and inert, have a high central surface brightness, and cover a wide range of luminosities (\cite[Driver \textit{et al.} 2006]{driver2006}). ETGs are classed as smooth, highly concentrated, and spheroidal systems with a lack of spiral arms. Their luminosity profiles tend to follow the de Vaucouleurs I(r)$\sim$r$^{1/4}$ law  or a more general Sersic law distribution (\cite[D'Onofrio \textit{et al.} 2011]{donofrio2011}).

ETGs are thought to be quiescent at zero redshift, leading to the assumption that they are devoid of both cold gas and dust (\cite[Bregman, Hogg \& Roberts 1992]{bregman1992}). Recent detections of ETGs in infrared and sub-mm regimes have revealed largely unexpected amount of cold gas and dust (eg. \cite[Leeuw \textit{et al.} (2004)]{leeuw2004}). Our interests lie in exploring this dusty presence, with the aims of gaining further understanding about the processes which form ETGs.

\begin{figure}
\begin{center}
 \includegraphics[width=5.0in]{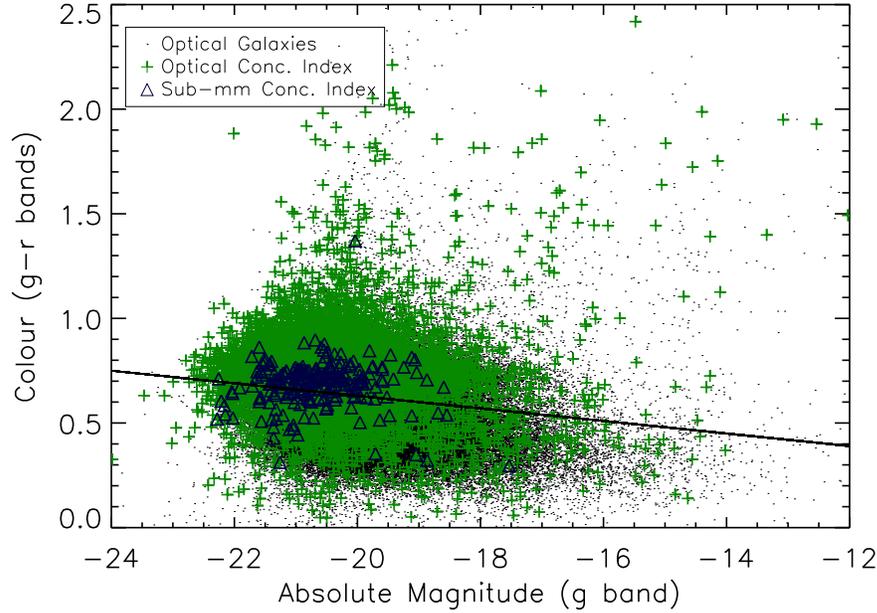} 
 \caption{Optical Colour-Magnitude Diagram for our samples. The highly concentrated (C$\ge$2.86) optical GAMA sources are highlighted by green plus signs above a background of all the GAMA sources (black dots). On top of this is the sub-mm selected sources from the combined GAMA/H-ATLAS catalogues, shown as dark blue open triangles. The Red Sequence (RS) line (\cite[Bernardi \textit{et al.} 2010]{Bernardietal10}) is shown to indicate the regime above which we expect RS galaxies to sit. This plot shows that our sample of ETGs could not have been selected with a direct colour cut, as both blue and red ETGs are showing up. This is confirmed through visual classifications of the samples.}
   \label{fig1}
\end{center}
\end{figure}

\begin{figure}
 \hspace*{-1cm}
 \includegraphics[width=0.6\textwidth]{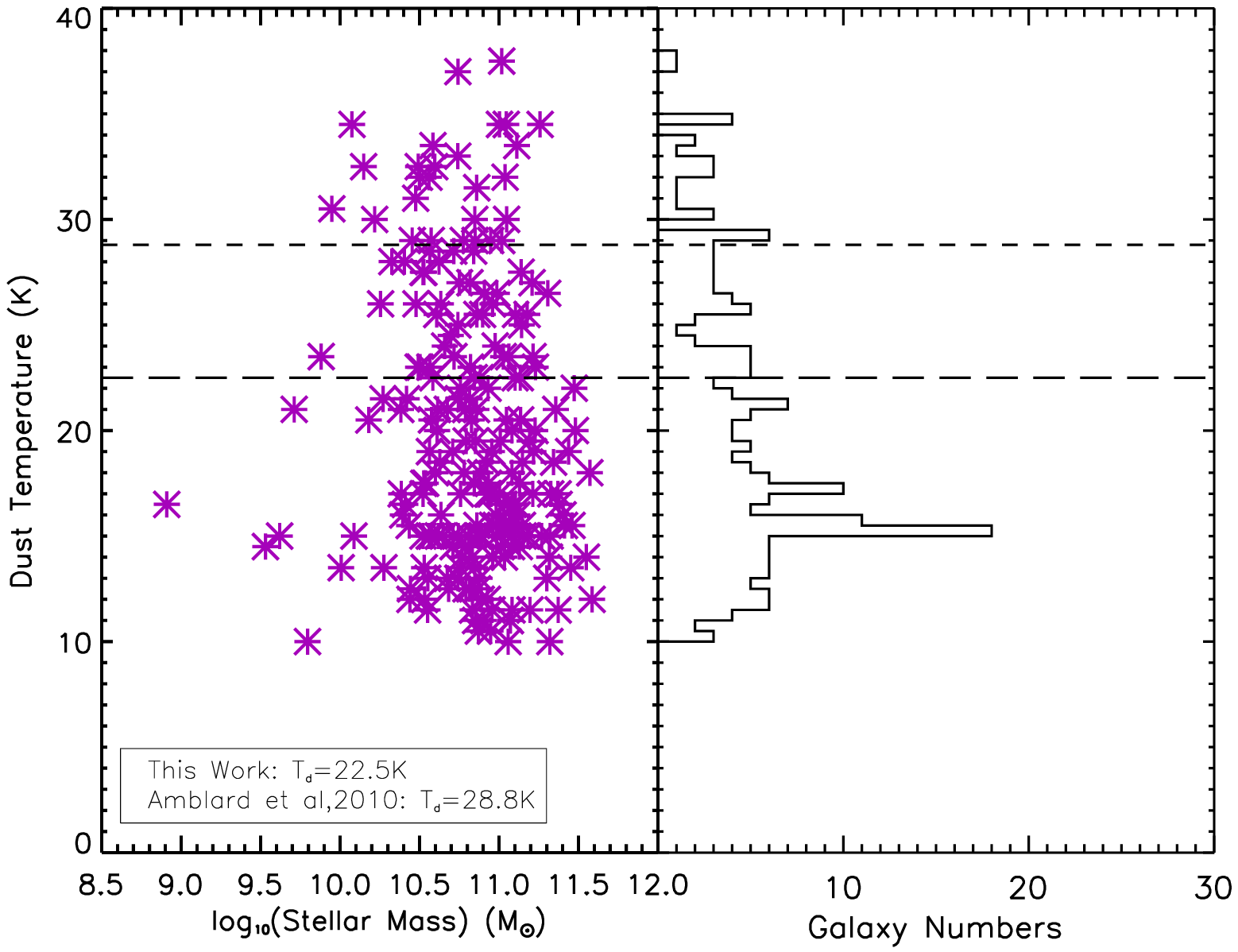} 
 \hspace*{-1cm}
\includegraphics[width=0.6\textwidth]{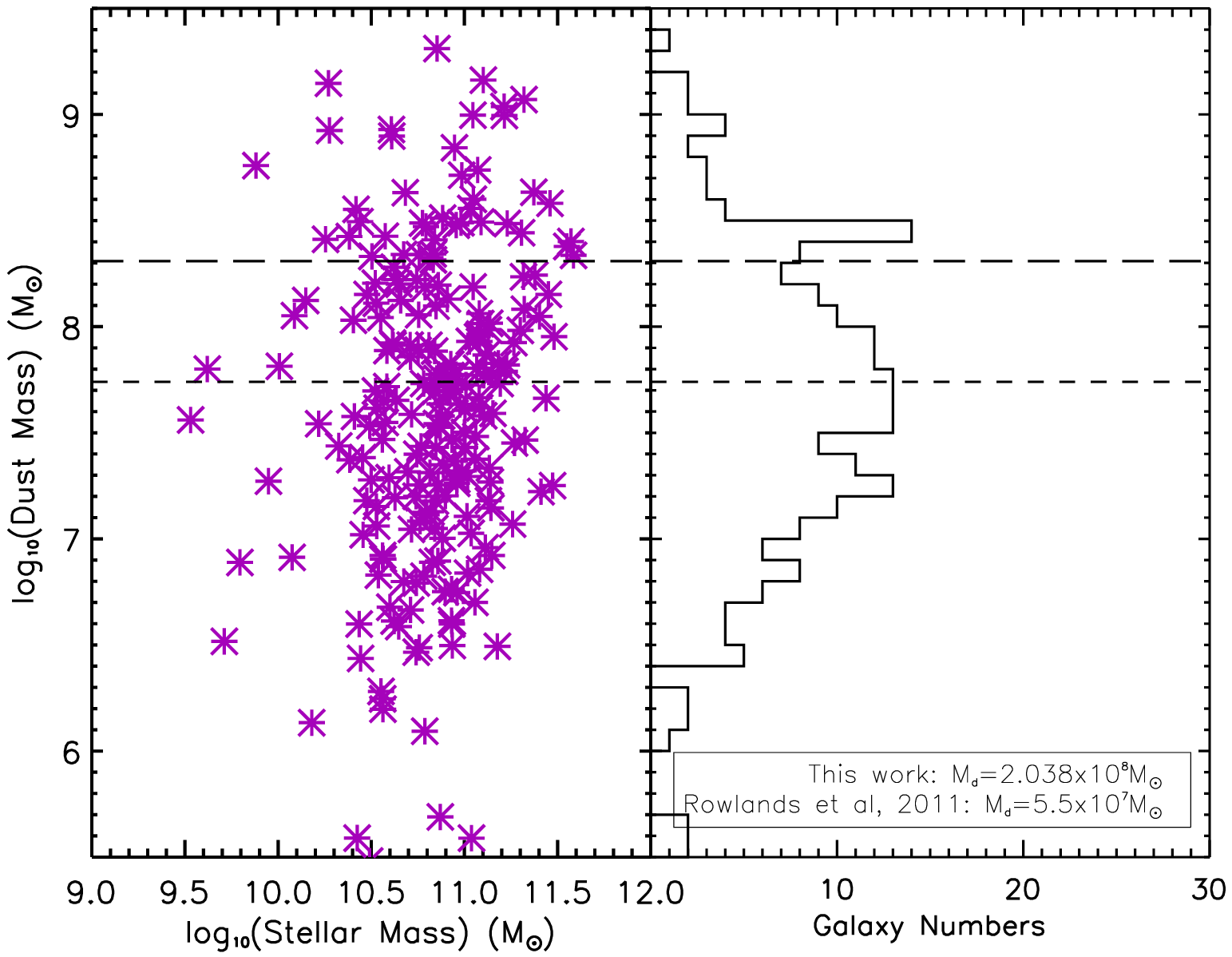}
\caption{[Left] The distribution of isothermal dust temperatures for the ETG sample with concentration index $\ge$ 2.86. Our mean temperature is the long dashed line and the short dashed line is the mean temperature from  \cite[Amblard \textit{et al.} (2010)]{Amblardetal10}  (for the H-ATLAS Science Demonstration Phase (SDP) sample). [Right] The distribution of dust masses fit by modified Planck functions, emphasising our mean dust mass (long dashed line) and that of the SDP ETG sample of \cite[Rowlands \textit{et al.} (2011)]{Rowlandsetal11} (short dashed line).}
   \label{fig2}

\end{figure}

\section{Selection Techniques}

With the aim of creating a highly robust sample of sub-mm selected ETGs, we experimented with multiple proxies for morphology. Proxies tested were optical colour (eg. \cite[Kaviraj \textit{et al.} 2010]{kaviraj2010}), r band Sersic index (eg. \cite[Leeuw \textit{et al.} 2008]{leeuw2008}) and r band Concentration index (eg. \cite[Blanton \etal\ 2003]{blanton2010)}. We used Sersic index values from the GAMA single Sersic fits to the objects, using GALFIT3 (\cite[Kelvin \textit{et al.} submitted]{Kelvininprep}). The Concentration values are the ratio of SDSS Petrosian 90 radius to Petrosian 50 radius for each galaxy. We combined the GAMA I multi-wavelength survey with Herschel-ATLAS Phase 1 far infrared and sub-mm data to create a catalogue of sub-mm detected ETGs.

We examined the behaviour of the three proxies for both our matched multi-wavelength sample and an optical morphologically classified SDSS sample from \cite[Nair \& Abraham (2010)]{NairAbe2010}. From this we concluded that a straight colour cut is not a viable approach to take, as ETGs tend to be red, but not all red galaxies tend to be ETGs. For example, many red spirals were present in such a cut. Also, as our sub-mm approach picks out dusty galaxies, we may expect some blue ETGs to be present. A Sersic index cut at $\ge$3.5 seemed reasonable, but a concentration index cut of $\ge$2.86 gave us a sample which matched best with the optical visually classified ETGs in the Nair $\&$ Abraham control sample.

We have applied our concentration index criterion, with additional criteria for redshift, ellipticity and apparent effective radius, to our GAMA/H-ATLAS data. With the addition of visual classifications to root out any remaining late-type galaxies, this resulted in a condensed sample of 239 lenticular and elliptical galaxies, shown as open triangles in Fig.\,\ref{fig1}. These sub-mm detected ETGs mostly have intermediate absolute magnitudes (-19.5$\sim$M$_{g}\sim$-21.5).

\section{Sub-mm Single Temperature Fits}
Modified Planck functions were fit to the sub-mm regime of our sample's observed data. The resultant cold dust temperatures and masses are displayed in Fig.\,\ref{fig2}. We find a mean dust temperature of 22.5$^{+10}_{-6}$K (for $\beta=$2.0) and a mean dust mass of 2.038$\times$10$^{8}$M$_{\odot}$. Fig.\,\ref{fig2}  shows that the sub-mm sample contains some ETGs with high dust masses and mainly low dust temperatures compared to these references. The low dust temperatures will need to be checked with additional flux points, including PACS data, covering the peak of the cool emission. Unlike the H-ATLAS galaxies which have been studied up to this point, our sample exhibits only intermediate stellar masses, not high stellar masses.

\end{document}